# Classical Linear Harmonic Oscillators to Describe Thermodynamic Properties of Quantum Linear Harmonic Oscillators and Solids

**Ikhtier Umirzakov** [1,2,3*]

1. Institute of Thermophysics, Lavrentev prospect, 1, 630090 Novosibirsk, Russia
2. Institute of Chemical Kinetics and Combustion, Institutskaya street, 3, 630090 Novosibirsk, Russia
3. Butlerov Scientific Foundation, Bondarenko street, 33-34, 420066, Kazan, Tatarstan Republic, Russia
* Correspondence: cluster125@gmail.com

**Abstract:** As known all physical properties of solids are described well by the system of quantum linear harmonic oscillators. It is shown in the present paper that the system consisting of classical linear harmonic oscillators having temperature dependent masses and/or frequencies has the same partition function as the system consisting of quantum linear harmonic oscillators having temperature independent masses and frequencies while the means of the square displacements of the positions of the oscillators from their mean positions for the system consisting of classical linear harmonic oscillators having: the temperature dependent masses; temperature dependent frequencies;  and temperature dependent masses and frequencies differ from each other and from that of the system consisting of quantum linear harmonic oscillators, and hence, the system consisting of classical linear harmonic oscillators describes well the thermodynamic properties of the system consisting of quantum linear harmonic oscillators and solids.

**1. Introduction**

The physical properties of solids are described well by the system consisting of quantum linear harmonic oscillators [1]. If the frequency spectrum of the system of linear harmonic oscillators is known then all thermodynamic properties of the system can be defined [1, 2]. Usually the Debye and Einstein spectra are used [1-5].

We show in the present paper that the system consisting of independent classical linear harmonic oscillators having temperature dependent masses or (and) frequencies has the same partition function as the system consisting of independent quantum linear harmonic oscillators having temperature independent masses and frequencies while the means of the square displacements of the positions of the oscillators from their mean positions of the system consisting of classical linear harmonic oscillators having: the temperature dependent masses; temperature dependent frequencies;  and temperature dependent masses and frequencies differ from each other and from that of the system of consisting quantum linear harmonic oscillators, and hence, the system consisting of classical linear harmonic oscillators describes well the thermodynamic properties of the system consisting of quantum linear harmonic oscillators,  and, hence, solids.

## 2. Quantum One-Dimensional Linear Harmonic Oscillator

The Hamilton operator $\hat{H}_{qu}$ of the one-dimensional linear harmonic oscillator is given by

$$\hat{H}_{qu}(p,x) = \frac{\hat{p}^2}{2m} + \frac{m\omega^2 x^2}{2}, \qquad (1)$$

where $\omega = const$ is the frequency of the oscillator, $m = const$ is the mass of the oscillator, $x$ is the position of the oscillator, $\hat{p} = -i\hbar \cdot \partial/\partial x$ is the momentum operator, $i$ is the imaginary unit and $\hbar$ is the Planck constant [5,6].

The partition function $Z_{qu}$ of the quantum one-dimensional linear harmonic oscillator is defined by

$$Z_{qu}\left(\frac{\hbar\omega}{2kT}\right) = \left[\exp\left(\frac{\hbar\omega}{2kT}\right) - \exp\left(-\frac{\hbar\omega}{2kT}\right)\right]^{-1}, \qquad (2)$$

where $k$ is the Boltzmann constant and $T$ is the absolute temperature [1,5,6].

The mean of the square displacements $\sigma_{qu}^2$ of the position of the quantum one-dimensional harmonic oscillator from its mean position is given by the following relation [6]:

$$\frac{m\omega\sigma_{qu}^2}{\hbar} = \frac{1}{2} \cdot \frac{\exp(\hbar\omega/kT)+1}{\exp(\hbar\omega/kT)-1}. \qquad (3)$$

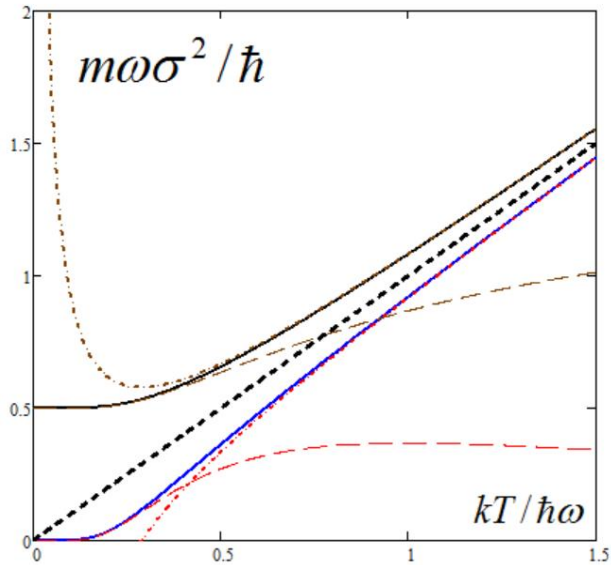

Fig. 1. Dependence of reduced square displacements $m\omega\sigma^2/\hbar$ of the position of the one-dimensional harmonic oscillator from its mean position on reduced temperature $kT/\hbar\omega$. Black solid upper line corresponds to Eq. 3, brown dashed-dotted upper line - to Eq. 4, brown dashed upper line - to Eq. 5. Blue solid lower line corresponds to Eq. 17, red dashed-dotted lower line - to Eq. 19, red dashed lower line - to Eq. 20. Black dashed straight line corresponds to Eqs. 21 and 28.

One can establish from Eq. 3 and Fig. 1 that

$$\frac{m\omega\sigma_{qu}^2}{\hbar} \approx \frac{kT}{\hbar\omega} \cdot \left[1 + \frac{1}{12}\left(\frac{\hbar\omega}{kT}\right)^2\right] \qquad (4)$$

for $kT/\hbar\omega > 1/2$, and

$$\frac{m\omega\sigma_{qu}^2}{\hbar} \approx \frac{1}{2}\left[1+2\exp\left(-\frac{\hbar\omega}{kT}\right)\right] \tag{5}$$

for $kT/\hbar\omega < 1/2$.

Besides it is easy to see using Eq. 3 that $\sigma_{qu}^2\big|_{T\to 0} = \hbar/2m\omega = const$ and $\sigma_{qu}^2\big|_{T\to\infty} = kT/m\omega^2$.

## 3. Classical One-Dimensional Linear Harmonic Oscillator Having Temperature Dependent Parameters

**Model 1.** Let as consider the classical one-dimensional linear harmonic oscillator having a temperature dependent frequency $\Omega_{cl}(T)$ and the following Hamilton function $H_{cl}(p,x)$

$$H_{cl}(p,x) = \frac{p^2}{2m} + \frac{m\Omega_{cl}^2(T)x^2}{2}, \tag{6}$$

where $p$ is the momentum of the oscillator. The partition function $Z_{cl}$ of the classical oscillator, defined from the following relation [2]

$$Z = \frac{1}{2\pi\hbar}\int_{-\infty}^{\infty}dx\int_{-\infty}^{\infty}\exp\left[-\frac{H(p,x)}{kT}\right]dp, \tag{7}$$

is given by the following relation

$$Z_{cl} = kT/\hbar\Omega_{cl}(T). \tag{8}$$

Using the equality $Z_{cl} = Z_{qu}$, Eqs. 2 and 8 we can conclude that the partition functions of the one-dimensional quantum and classical linear harmonic oscillators are equal to each other if the frequency of the classical oscillator is defined from the following relation

$$\frac{\Omega_{cl}(T)}{\omega} = \frac{kT}{\hbar\omega}\cdot\left[\exp\left(\frac{\hbar\omega}{2kT}\right) - \exp\left(-\frac{\hbar\omega}{2kT}\right)\right]. \tag{9}$$

One can establish from Eq. 9 and Fig. 2 that

$$\frac{\Omega_{cl}(T)}{\omega} \approx 1 + \frac{1}{24}\left(\frac{\hbar\omega}{kT}\right)^2 \tag{10}$$

for $kT/\hbar\omega > 1/2$, and

$$\frac{\Omega_{cl}(T)}{\omega} \approx \frac{kT}{\hbar\omega}\cdot\exp\left(\frac{\hbar\omega}{2kT}\right) \tag{11}$$

for $kT/\hbar\omega < 1/4$.

The density $\rho_{cl}(x)$ of the distribution of the position of the classical one-dimensional linear harmonic oscillator is given by

$$\rho_{cl}(x) = \sqrt{\frac{m\Omega_{cl}^2(T)}{2\pi kT}}\exp\left[-\frac{m\Omega_{cl}^2(T)x^2}{2kT}\right]. \tag{12}$$

We have from Eqs. 9 and 12

$$\rho_{cl}(x) = \sqrt{\frac{m\omega}{2\pi\hbar}} \cdot \frac{\exp\left(\frac{\hbar\omega}{2kT}\right) - \exp\left(-\frac{\hbar\omega}{2kT}\right)}{\sqrt{\frac{\hbar\omega}{kT}}} \exp\left\{-\frac{mx^2\omega}{2\hbar(\hbar\omega/kT)}\left[\exp\left(\frac{\hbar\omega}{2kT}\right) - \exp\left(-\frac{\hbar\omega}{2kT}\right)\right]^2\right\}. \quad (13)$$

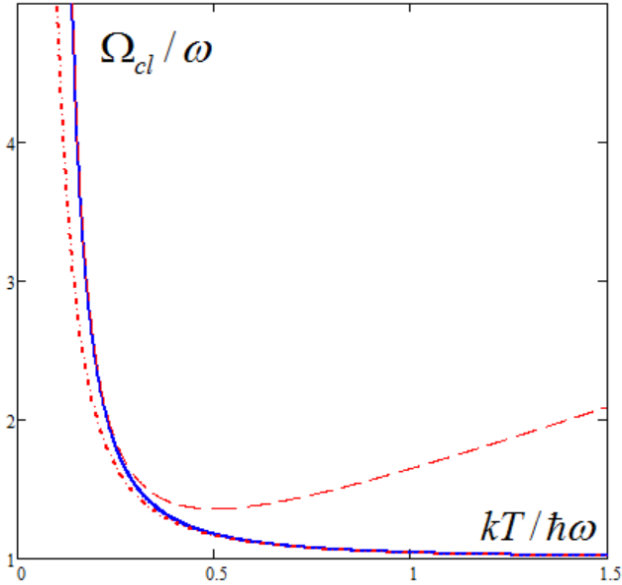

Fig. 2. Dependence of reduced frequency $\Omega_{cl}/\omega$ of classical one-dimensional harmonic oscillator on reduced temperature $kT/\hbar\omega$. Blue solid line corresponds to Eq. 9, red dashed-dotted line - to Eq. 10, red dashed line - to Eq. 11.

The mean $<\varphi(x)>_{cl}$ of an arbitrary function $\varphi(x)$ of the position $x$ is defined by

$$<\varphi(x)>_{cl} \equiv \int_{-\infty}^{\infty} \varphi(x)\rho_{cl}(x)dx. \quad (14)$$

Using Eq. 14 it is easy to obtain

$$<x>_{cl} = 0, \quad (15)$$

and

$$\sigma_{cl}^2 = \frac{kT}{m\Omega_{cl}^2(T)} = \frac{\hbar^2}{mkT}\left[\exp\left(\frac{\hbar\omega}{2kT}\right) - \exp\left(-\frac{\hbar\omega}{2kT}\right)\right]^{-2} \quad (16)$$

or

$$\frac{m\omega\sigma_{cl}^2}{\hbar} = \frac{\hbar\omega}{kT}\left[\exp\left(\frac{\hbar\omega}{2kT}\right) - \exp\left(-\frac{\hbar\omega}{2kT}\right)\right]^{-2} \quad (17)$$

for the mean of the square displacements $\sigma_{cl}^2$ of the position of the classical one-dimensional harmonic oscillator from its mean position which is defined by

$$\sigma_{cl}^2 \equiv <x^2>_{cl} - <x>_{cl}^2. \quad (18)$$

We can conclude using Eq. 17 and Fig. 1 that

$$\frac{m\omega\sigma_{cl}^2}{\hbar} \approx \frac{kT}{\hbar\omega}\left[1 - \frac{1}{12}\left(\frac{\hbar\omega}{kT}\right)^2\right] \qquad (19)$$

for $kT/\hbar\omega > 1/2$, and

$$\frac{m\omega\sigma_{cl}^2}{\hbar} \approx \frac{\hbar\omega}{kT}\exp\left(-\frac{\hbar\omega}{kT}\right) \qquad (20)$$

for $kT/\hbar\omega < 1/4$.

We can conclude from Eqs. 17, 19 and 20 and Fig. 1 that the mean square displacements $\sigma_{cl}^2$ of the position of the oscillator increases with increasing temperature, and $\sigma_{cl}^2$ is extremely small at low temperatures.

One can conclude from Eqs. 9-11 and Fig. 2 that the frequency $\Omega_{cl}(T)$ of the oscillator decreases with increasing temperature, the frequency is extremely large at low temperatures, and $\Omega_{cl}(T)|_{T\to\infty} = \omega$.

We obtain from Eqs. 3 and 17

$$\left.\frac{m\omega\sigma_{qu}^2}{\hbar}\right|_{T\to\infty} = \left.\frac{m\omega\sigma_{cl}^2}{\hbar}\right|_{T\to\infty} = \frac{kT}{\hbar\omega}. \qquad (21)$$

Having known partition function one can define all thermodynamic properties of the oscillator [2]. So we showed that the classical one-dimensional linear harmonic oscillator having the temperature dependent frequency defined by Eq. 9 describes the thermodynamic properties of the quantum linear harmonic oscillator.

**Model 2.** Let us consider classical one-dimensional linear harmonic oscillator having a temperature dependent mass $\mu(T)$ and the following Hamilton function

$$H_2(p,x) = \frac{p^2}{2\mu(T)} + \frac{m\omega^2 x^2}{2}. \qquad (22)$$

The partition function $Z_2$ of this oscillator is equal to

$$Z_2 = g(T)\frac{kT}{\hbar\omega}, \qquad (23)$$

where $g(T) = \sqrt{\mu(T)/m}$. It is easy to see that $Z_2 = Z_{qu}$ if

$$g(T) \equiv \frac{\hbar\omega}{kT}\left[\exp\left(\frac{\hbar\omega}{2kT}\right) - \exp\left(-\frac{\hbar\omega}{2kT}\right)\right]^{-1}. \qquad (24)$$

So we showed that the classical one-dimensional linear harmonic oscillator having the temperature dependent mass defined by

$$\mu(T) = m\left(\frac{\hbar\omega}{kT}\right)^2\left[\exp\left(\frac{\hbar\omega}{2kT}\right) - \exp\left(-\frac{\hbar\omega}{2kT}\right)\right]^{-2} \qquad (25)$$

can describe the thermodynamic properties of the quantum linear harmonic oscillator.

Using Eq. 25 and Fig. 3 we established that

$$\frac{\mu(T)}{m} \approx 1 - \frac{1}{12}\left(\frac{\hbar\omega}{kT}\right)^2 \qquad (26)$$

for $kT/\hbar\omega > 1$, and

$$\frac{\mu(T)}{m} \approx \left(\frac{\hbar\omega}{kT}\right)^2 \exp\left(-\frac{\hbar\omega}{kT}\right) \qquad (27)$$

for $kT/\hbar\omega < 1/4$.

One can conclude from Eqs. 25-27 and Fig. 3 that the mass $\mu(T)$ of the oscillator increases with increasing temperature, the mass is extremely small at low temperatures and $\mu(T)|_{T\to\infty} = m$.

So we showed that the classical one-dimensional linear harmonic oscillator having the temperature dependent mass defined by Eq. 25 describes the thermodynamic properties of the quantum linear harmonic oscillator.

The mean of the square displacements $\sigma_2^2$ of the position of the classical one-dimensional linear harmonic oscillator from its mean position is given by

$$\sigma_2^2 = \frac{kT}{m\omega^2}. \qquad (28)$$

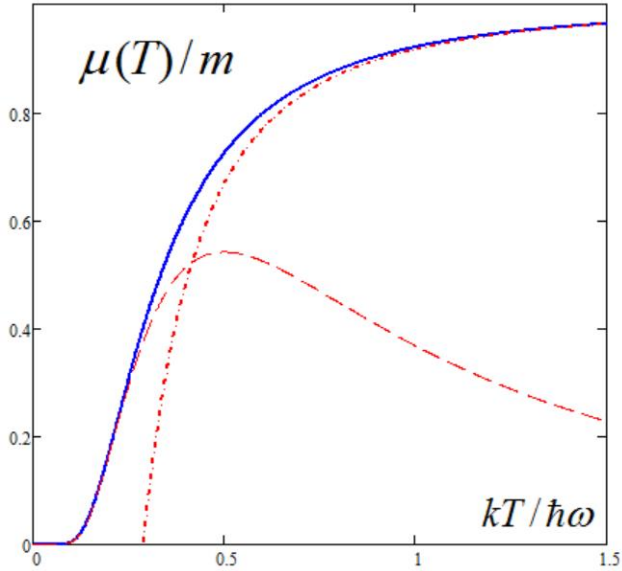

Fig. 3. Dependence of reduced mass $\mu/m$ of classical one-dimensional harmonic oscillator on reduced temperature $kT/\hbar\omega$. Blue solid line corresponds to Eq. 25, red dashed-dotted line - to Eq. 26, red dashed line - to Eq. 27.

**Model 3.** It is easy to see that the partition function of the classical one-dimensional linear harmonic oscillator having the temperature dependent mass $\bar{\mu}(T)$ and frequency $\Omega_{cl}(T)$ given by Eq. 9 and the following Hamiltonian

$$H_3(p,x) = \frac{p^2}{2\bar{\mu}(T)} + \frac{\bar{\mu}(T)\Omega_{cl}^2(T)x^2}{2}, \qquad (29)$$

is equal to the partition function of the quantum one-dimensional harmonic oscillator (see Eq. 2). Here $\bar{\mu}(T)$ is an arbitrary positive function of temperature, and, particularly, $\bar{\mu}(T) = \mu(T)$, where $\mu(T)$ is defined by Eq. 25.

So we proved that the classical one-dimensional linear harmonic oscillator having the temperature dependent mass and frequency can describe the thermodynamic properties of the quantum linear harmonic oscillator.

The mean of the square displacements $\sigma_3^2$ of the position of the classical one-dimensional linear harmonic oscillator from its mean position is given by

$$\sigma_3^2 = \frac{kT}{\bar{\mu}(T)\Omega_{cl}^2(T)}. \tag{30}$$

One can see from Eqs. 16, 28 and 30 that the means of the square displacements of the position of the oscillator from its mean position for the classical one-dimensional linear harmonic oscillator having: the temperature dependent mass; temperature dependent frequency; and the temperature dependent mass and frequency differ from each other.

The comparison of Eq. 3 with Eqs. 16, 28 and 30 shows that the means of the square displacements of the position of the oscillator from its mean position for the classical one-dimensional oscillator having: the temperature dependent mass; temperature dependent frequency; and the temperature dependent mass and frequency differ from that of the quantum one-dimensional linear harmonic oscillator.

So we showed that the classical one-dimensional linear harmonic oscillator having the temperature dependent mass and/or frequency has the same partition function as the quantum linear harmonic oscillator having the temperature independent mass and frequency while the means of the square displacements of the position of the oscillator from its mean position for the classical linear harmonic oscillator having: the temperature dependent mass; the temperature dependent frequency; and the temperature dependent mass and frequency differ from each other and from that of the system of quantum one-dimensional linear harmonic oscillator. Hence, the classical one-dimensional linear harmonic oscillator having temperature dependent mass or (and) frequency can describe all thermodynamic properties of the quantum one-dimensional linear harmonic oscillator.

## 4. The System of Independent Quantum One-Dimensional Linear Harmonic Oscillators

Let us consider the system consisting of $N$ atoms or molecules interacting with each other. The system has $3N-6$ independent linear harmonic vibrational modes at low temperatures [1].

The quantum Hamilton operator $\hat{H}$ of the system consisting of independent one-dimensional linear harmonic oscillators is given by

$$\hat{H} = \sum_{j=1}^{3N-6}\left(\frac{\hat{p}_j^2}{2m_j} + \frac{m_j\omega_j^2 x_j^2}{2}\right), \tag{31}$$

where $j$ is the order number of the oscillator (vibrational mode), $x_j$, $\hat{p}_j = -i\hbar\cdot\partial/\partial x_j$, $\omega_j$ and $m_j$ are the position, momentum operator, frequency and mass of the $j$-th oscillator, respectively [1-6]. The partition function of the system consisting of independent linear harmonic oscillators is given by

$$Z = \prod_{j=1}^{3N-6} Z_{qu}\left(\frac{\hbar\omega_j}{2kT}\right), \tag{32}$$

where $Z_{qu}(\hbar\omega_j/2kT)$ is given by Eq. 2 [1-6].

Using Eq. 3 we obtain

$$\bar{\sigma}_{qu}^2 = \frac{\hbar}{2(3N-6)} \sum_{j=1}^{3N-6} \left[ \frac{1}{m_j\omega_j} \cdot \frac{\exp(\hbar\omega_j/kT)+1}{\exp(\hbar\omega_j/kT)-1} \right] \tag{33}$$

for the mean of the square displacements $\bar{\sigma}_{qu}^2$ of the positions of the quantum one-dimensional linear harmonic oscillators from their mean positions which is defined by $\bar{\sigma}^2 = \frac{1}{3N-6} \sum_{j=1}^{3N-6} \bar{\sigma}_j^2$, where $\bar{\sigma}_j^2$ is the mean of the square displacements of the position of the $j$-th one-dimensional linear harmonic oscillator from its mean position.

## 5. The System of Classical Independent One-Dimensional Linear Harmonic Oscillators Having Temperature Dependent Parameters

**Model A.** Let us consider the system consisting of classical independent one-dimensional linear harmonic oscillators having a temperature dependent frequencies $\Omega_{cl,j}(T)$ and the following Hamilton function

$$H_1 = \sum_{j=1}^{3N-6} H_j(p_j, x_j), \tag{34}$$

where

$$H_j(p_j, x_j) = \frac{p_j^2}{2m_j} + \frac{m_j \Omega_{cl,j}^2(T) x_j^2}{2}, \tag{35}$$

$p_j$ is the momentum of $j$-oscillator and $\Omega_{cl,j}(T)$ is given by

$$\frac{\Omega_{cl,j}(T)}{\omega_j} = \frac{kT}{\hbar\omega_j} \cdot \left[ \exp\left(\frac{\hbar\omega_j}{2kT}\right) - \exp\left(-\frac{\hbar\omega_j}{2kT}\right) \right]. \tag{36}$$

Using Eq. 34-36 and the following relation for the partition function of the system consisting of classical independent linear harmonic oscillators [2]

$$\bar{Z} = \prod_{j=1}^{3N-6} \frac{1}{2\pi\hbar} \int_{-\infty}^{\infty} dx_j \int_{-\infty}^{\infty} \exp\left[-\frac{H_j(p_j, x_j)}{kT}\right] dp_j, \tag{37}$$

that $\bar{Z} = Z$, where $Z$ is given by Eq. 32.

Having known partition function one can define all thermodynamic properties of the system consisting of independent linear harmonic oscillators [2].

So we proved that the system consisting of classical independent one-dimensional linear harmonic oscillators having the temperature dependent frequencies defined by Eqs. 36 describes the thermodynamic properties of the system consisting of quantum independent linear harmonic oscillators.

Using Eq. 16 we established the following relation

$$\overline{\sigma}_{cl}^2 = \frac{\hbar^2}{kT(3N-6)} \sum_{j=1}^{3N-6} \frac{1}{m_j} \left[ \exp\left(\frac{\hbar\omega_j}{2kT}\right) - \exp\left(-\frac{\hbar\omega_j}{2kT}\right) \right]^{-2} \qquad (38)$$

for the mean of the square displacements $\overline{\sigma}_{cl}^2$ of the positions of the classical one-dimensional linear harmonic oscillators from their mean positions.

**Model B.** This model represents the system consisting of classical independent one-dimensional linear harmonic oscillators having a temperature dependent masses $\mu_j(T)$ and the following Hamilton function

$$\overline{H}_2 = \sum_{j=1}^{3N-6} H_{2j}(p_j, x_j), \qquad (39)$$

where

$$H_{2j}(p_j, x_j) = \frac{p_j^2}{2\mu_j(T)} + \frac{m_j \omega_j^2 x_j^2}{2}, \qquad (40)$$

$$\mu_j(T) = m_j \left(\frac{\hbar\omega_j}{kT}\right)^2 \left[ \exp\left(\frac{\hbar\omega_j}{2kT}\right) - \exp\left(-\frac{\hbar\omega_j}{2kT}\right) \right]^{-2}. \qquad (41)$$

It is easy to establish using Eqs. 37 and 39-41 that the partition function of the system consisting of the classical independent one-dimensional linear harmonic oscillators having the Hamilton function given by Eq. 39-41 is equal to that of the system consisting of the quantum independent one-dimensional linear harmonic oscillators given by Eq. 32.

So we showed that the system consisting of independent classical one-dimensional linear harmonic oscillators having the temperature dependent masses defined by Eqs. 41 describes the thermodynamic properties of the system consisting of independent quantum linear harmonic oscillators.

Using Eq. 28 we established the following relation

$$\overline{\sigma}_2^2 = \frac{kT}{3N-6} \sum_{j=1}^{3N-6} \frac{1}{m_j \omega_j^2} \qquad (42)$$

for the mean of the square displacements $\overline{\sigma}_2^2$ of the positions of the classical one-dimensional linear harmonic oscillators from their mean positions.

**Model C.** This model represents the system consisting of classical independent one-dimensional linear harmonic oscillators having a temperature dependent masses $\overline{\mu}_j(T)$ and frequencies $\Omega_{cl,j}(T)$, which has the following Hamilton function

$$\overline{H}_3 = \sum_{j=1}^{3N-6} H_{3j}(p_j, x_j) \qquad (43)$$

where

$$H_{3j}(p_j, x_j) = \frac{p_j^2}{2\overline{\mu}_j(T)} + \frac{\overline{\mu}_j(T)\Omega_{cl,j}^2(T) x_j^2}{2}, \qquad (44)$$

$\Omega_{cl,j}(T)$ is given by Eq. 36 and $\overline{\mu}_j(T)$ is an arbitrary positive function of temperature.

Using Eqs. 36, 37, 43 and 44 one can easily show that the partition function of the system consisting of the classical independent one-dimensional linear harmonic oscillators having the Hamilton function given by Eqs. 36, 43 and 44 is equal to that of the system consisting of the quantum independent one-dimensional linear harmonic oscillators given by Eq. 32.

So we showed that the system consisting of independent classical one-dimensional linear harmonic oscillators having the arbitrary temperature dependent masses and frequencies defined by Eqs. 36 describes the thermodynamic properties of the system consisting of independent quantum linear harmonic oscillators.

Using Eq. 30 we established the following relation

$$\overline{\sigma}_3^2 = \frac{kT}{3N-6} \sum_{j=1}^{3N-6} \frac{1}{\overline{\mu}_j(T)\Omega_{cl,j}^2(T)} \tag{45}$$

for the mean of the square displacements $\overline{\sigma}_3^2$ of the positions of the classical one-dimensional linear harmonic oscillators from their mean positions.

**Model D.** This model consists of $N_1$ classical independent one-dimensional linear harmonic oscillators with Hamilton function given by Eqs. 34-36, $N_2$ classical independent one-dimensional linear harmonic oscillators with Hamilton function given by Eqs. 39-41, and $3N-6-N_1-N_2$ classical independent one-dimensional linear harmonic oscillators with Hamilton function given by Eqs. 43-44. Here $0 \leq N_1 \leq 3N-6$, $0 \leq N_2 \leq 3N-6$ and $0 < N_1 + N_2 \leq 3N-6$. This system of oscillators has the following Hamilton function

$$\overline{H} = \sum_{j=1}^{N_1} H_j(p_j, x_j) + \sum_{j=N_1+1}^{N_1+N_2} H_{2j}(p_j, x_j) + \sum_{j=N_1+N_2+1}^{3N-6} H_{3j}(p_j, x_j). \tag{46}$$

Using Eqs. 34-37, 39-41 and 43-44 one can easily show that the partition function of the system consisting of the classical independent one-dimensional linear harmonic oscillators having the Hamilton function given by Eq. 46 is equal to that of the system consisting of the quantum independent one-dimensional linear harmonic oscillators given by Eq. 32.

So we showed that the system consisting of independent classical one-dimensional linear harmonic oscillators having arbitrary temperature dependent masses and frequencies and the Hamilton function given by Eq. 46 can describe the thermodynamic properties of the system consisting of independent quantum linear harmonic oscillators.

Using Eqs. 16, 28 and 30 we obtain the following relation

$$\overline{\sigma}^2 = \frac{kT}{3N-6} \left\{ \sum_{j=1}^{N_1} \frac{\hbar^2}{m_j k^2 T^2} \left[ \exp\left(\frac{\hbar\omega_j}{2kT}\right) - \exp\left(-\frac{\hbar\omega_j}{2kT}\right) \right]^{-2} + \sum_{j=N_1+1}^{N_1+N_2} \frac{1}{m_j \omega_j^2} + \sum_{j=N_1+N_2+1}^{3N-6} \frac{\Omega_{cl,j}^{-2}(T)}{\overline{\mu}_j(T)} \right\} \tag{47}$$

for the mean of the square displacements $\overline{\sigma}^2$ of the positions of the classical one-dimensional linear harmonic oscillators from their mean positions.

One can see from Eqs. 38, 42, 45 and 47 that the means of the square displacements of the positions of the oscillators from their mean positions for the system consisting of the classical one-dimensional linear harmonic oscillators having: the temperature dependent masses; temperature dependent frequencies; and temperature dependent masses and frequencies differ from each other.

Comparing Eq. 33 with Eqs. 38, 42, 45 and 47 we can conclude that the means of the square displacements of the positions of the oscillators from their mean positions for the system consisting of the classical independent one-dimensional linear harmonic oscillators having: the temperature dependent masses; temperature dependent frequencies; and temperature dependent masses and frequencies differ from that of the system consisting of the independent quantum linear harmonic oscillators.

So we showed that the system consisting of classical independent one-dimensional linear harmonic oscillators having temperature dependent masses and/or frequencies can describe all thermodynamic properties of the system consisting of quantum independent one-dimensional linear harmonic oscillators. The physical properties of solids are described well by the system consisting of quantum linear harmonic oscillators [1-6]. Hence, the system consisting of classical independent one-dimensional linear harmonic oscillators having temperature dependent masses and/or frequencies can describe all thermodynamic properties solids.

## 6. Conclusions

So we showed that:

1. the classical one-dimensional linear harmonic oscillator having the temperature dependent mass or (and) frequency has the same partition function as the quantum linear harmonic oscillator having the temperature independent mass and frequency while the means of the square displacements of the position of the oscillator from its mean position for the classical linear harmonic oscillator having: the temperature dependent mass; temperature dependent frequency; and temperature dependent mass and frequency differ from each other and from that of the system of quantum one-dimensional linear harmonic oscillator. Hence, the classical one-dimensional linear harmonic oscillator having the temperature dependent mass and/or frequency can describe all thermodynamic properties of the quantum one-dimensional linear harmonic oscillator;

2. the system consisting of classical linear harmonic oscillators having the temperature dependent masses or (and) frequencies has the same partition function as the system consisting of quantum linear harmonic oscillators having the temperature independent masses and frequencies while the means of the square displacements of the positions of the oscillators from their mean positions for the system consisting of classical linear harmonic oscillators having: the temperature dependent masses; the temperature dependent frequencies; and the temperature dependent masses and frequencies differ from each other and from that of the system consisting of quantum linear harmonic oscillators, and hence, the system consisting of classical linear harmonic oscillators describes well the thermodynamic properties of the system consisting of quantum linear harmonic oscillators, and hence, solids.

**References**

1. Kittel, Ch. *Introduction to Solid State Physics*; John Wiley and Sons Inc.: New York, USA, 1975.
2. Landau, L. D.; Lifshitz, E. M. *Statistical physics*; Butterworth-Heinemann: Oxford, 2013.
3. Debye, P. Zur Theorie der spezifischen Wärmen. *Ann. Physik* **1912**, 344, 789.
4. Einstein, A. Die Plancksche Theorie der Strahlung und die Theorie der spezifischen Wärme. *Ann. Physik* **1907**, 22, 180.
5. Landau, L.D., Lifshitz E.M. *Quantum mechanics: non-relativistic theory*; Elsevier: USA, 2013.
6. Feynman, R.P. *Statistical mechanics*; W.A. Benjamin: New York, USA, 1972.